\documentclass[10pt]{article}
\usepackage{ijgmmp-submission}
\newcommand{\Lm}{\mathcal{L}_m}
\newcommand{\Pbar}{\mathcal{P}}
\newcommand{\Rc}{R_c}
\newcommand{\dd}{\mathrm{d}}
\newcommand{\ee}{\mathrm{e}}
\newcommand{\Sch}{\mathrm{Sch}}
\newcommand{\ADM}{\mathrm{ADM}}

\newcommand{\XiB}{0.7768422203}
\newcommand{\HB}{0.541602901}
\newcommand{\RhoB}{-0.01496347214}

\newcommand{\FoldXP}{0.0297486727}
\newcommand{\FoldUP}{0.6142948408}
\newcommand{\FoldXR}{0.6214337869}
\newcommand{\FoldUR}{0.6780445951}
\newcommand{\AlphaCrit}{0.8795738643}
\newcommand{\ChiCrit}{-0.3864269455}
\newcommand{\UCrit}{0.6135730545}
\newcommand{\LBase}{2.402803104}
\newcommand{\LFamilyB}{1.490227195}

\hypersetup{pdftitle={C-infinity Compact-Support Wormholes with Exact Schwarzschild Exterior in Trace-Coupled Gravity},pdfauthor={Mushtaq Ahmad, M. Farasat Shamir, Ahdab K. Althukair},pdfkeywords={Traversable wormholes; compact support; trace-coupled gravity; Schwarzschild exterior; algebraic reconstruction}}
\begin{document}
\thispagestyle{plain}
\begin{center}
{\fontsize{10}{13}\selectfont\bfseries
$C^\infty$ Compact-Support Wormholes with Exact Schwarzschild Exterior in Trace-Coupled Gravity\par}
\vspace{13pt}
{\fontsize{8}{10}\selectfont Mushtaq Ahmad\textsuperscript{1,2,*}\par}
{\fontsize{8}{10}\selectfont\itshape
\textsuperscript{1}National University of Computer and Emerging Sciences, Islamabad, Pakistan\par
\textsuperscript{2}Center for Theoretical Physics, Khazar University, 41 Mehseti Str., AZ1096 Baku, Azerbaijan\par
\href{mailto:mushtaq.sial@nu.edu.pk}{mushtaq.sial@nu.edu.pk}\par}
\vspace{8pt}
{\fontsize{8}{10}\selectfont M. Farasat Shamir\textsuperscript{3}\par}
{\fontsize{8}{10}\selectfont\itshape
\textsuperscript{3}School of Computing and Mathematical Sciences, University of Leicester, United Kingdom\par
\href{mailto:mfs24@leicester.ac.uk}{mfs24@leicester.ac.uk}\par}
\vspace{8pt}
{\fontsize{8}{10}\selectfont Ahdab K. Althukair\textsuperscript{4}\par}
{\fontsize{8}{10}\selectfont\itshape
\textsuperscript{4}Department of Physics, College of Sciences, Princess Nourah bint Abdulrahman University,
P.O. Box 84428, Riyadh 11671, Saudi Arabia\par
\href{mailto:akalthukair@pnu.edu.sa}{akalthukair@pnu.edu.sa}\par}
\end{center}
{\fontsize{8}{10}\selectfont\noindent\textsuperscript{*}Corresponding author.\par}
\vspace{10pt}
\begin{abstract}
We construct static, spherically symmetric wormholes in trace-coupled gravity with
$f(R,T)=R+\lambda T^2$. Smooth deformations of Schwarzschild vanish to all orders
outside a finite core, producing an exact Schwarzschild exterior without a thin
shell. Under two explicitly assumed anisotropic-fluid closures, the matter
reconstruction reduces to a cubic equation for $\chi=\lambda T/(4\pi)$. The vacuum
boundary selects a locally unique root; its simple continuation must remain
separate from the singular inversion surface. We distinguish root folds from
vanishing $1+\chi$ and terminate numerical continuation when the selected root
coalesces. Regular examples have either positive or negative throat density.
On every regular branch the physical radial null energy condition fails at the
throat; for the undeformed-redshift family it fails throughout the open core.
The nonvacuum sector has finite support, whereas the exterior retains the
Schwarzschild Weyl field and mass $M$. Reproducible phase maps, convergence tests,
and high-precision tail checks support these conditional existence results.
No dynamical stability or microphysical matter model is asserted.
\end{abstract}
\vspace{4pt}
{\fontsize{8}{10}\selectfont\noindent
\par}

\section{Introduction}
\label{sec:introduction}
Traversable wormholes provide a setting in which spacetime topology, exact
solutions, and local energy conditions can be studied together
\cite{MorrisThorne1988,Visser1989,VisserBook1995,HochbergVisser1998,HochbergMolinaParisVisser1999}.
For a static spherical throat in Einstein gravity, flare-out forces a negative
radial null-energy combination. The question is therefore often how to confine
the supporting nonvacuum source while retaining a smooth, horizon-free geometry.

Here we require a finite core outside which the metric is exactly Schwarzschild.
The distinction between smooth matching and surface layers is familiar from
the junction-condition and thin-shell literature
\cite{Darmois1927,Israel1966,PoissonVisser1995,Eiroa2008,MazharimousaviHalilsoy2014}.
We use explicit flat cutoff functions to make the metric and every radial
derivative coincide with Schwarzschild at the core boundary. The resulting
compact boundary carries no distributional stress tensor.

Our matter reconstruction is formulated in the quadratic trace model
\begin{equation}
 f(R,T)=R+\lambda T^2,
 \label{eq:intro_model}
\end{equation}
within the metric framework of $f(R,T)$ gravity \cite{Harko2011}.
Since $f_R=1$, the metric equations remain second order. Explicit trace dependence
nevertheless makes the source reconstruction nonlinear and dependent on the
assumed matter prescription. Related wormhole analyses illustrate this
dependence and the distinction between physical and effective energy conditions
\cite{BanerjeeTangphatiPradhan2023,MoraesSahoo2017,MoraesSahoo2018,ChandaDeyPaul2021,MoraesSahoo2019,MishraSharmaDubeyPradhan2020,KiroriwalKumarMauryaChaudhary2024}.
Wormholes with linear and quadratic trace terms were also studied in Ref.~\cite{MoraesSahooQuadratic2018}; we do not claim novelty for the quadratic trace term itself.
As in other effective-source descriptions \cite{LoboOliveira2009}, geometric
regularity alone does not guarantee a regular physical matter realization.

Two qualifications are essential. First, the anisotropic-fluid prescriptions
used here are phenomenological closures for the tensor $\Theta_{\mu\nu}$;
we do not derive a microscopic matter action. Second, the nonlinear trace
coupling requires the complete divergence relation, including trace-gradient
terms, when force balance is checked. The correction to the original
$f(R,T)$ conservation equation was discussed by Barrientos and Rubilar
\cite{BarrientosRubilar2014}. The interpretation of separable trace terms as
nonlinear matter contributions has also been debated
\cite{FisherCarlson2019,HarkoMoraes2020,FisherCarlson2020}.
Our conclusions concern the displayed closure systems rather than resolving
that broader interpretational issue.

The construction separates two problems. The geometry is fixed first by a
compact-support deformation. The physical density and pressures are then
obtained from a cubic equation for $\chi=\lambda T/(4\pi)$.
At the vacuum boundary, $\chi=0$ is a simple root and selects a locally unique
continuation. Global admissibility requires that this selected root remain
simple and that $1+\chi$ stay positive throughout the core. A collision with
another root can end the continuation before the inversion denominator
vanishes. Tracking these two mechanisms separately prevents substitution of
a disconnected real root after a fold.

The numerical examples exhibit full regular branches with positive and negative
throat densities, together with parameter values for which the simple branch
ends inside the core. The physical radial null energy condition remains violated
at the throat. In the family without a redshift deformation, it is violated
throughout the open core and approaches zero from below at the compact boundary.
The gain is exact finite support, without a claim that the exotic region is
parametrically narrow or that the resulting configurations are dynamically stable.

Section~\ref{sec:setup} specifies the equations and closures.
Section~\ref{sec:compact} constructs the geometry.
Section~\ref{sec:reconstruction} establishes the continuation criteria and
energy-condition relations. Section~\ref{sec:numerics} presents the numerical
examples, Section~\ref{sec:regularity} discusses invariant regularity, and
Section~\ref{sec:reproducibility} gives reproducibility and convergence details.
Section~\ref{sec:conclusion} states the resulting limits of the construction.
\section{Field Equations and Wormhole System}
\label{sec:setup}

We use geometrized units, $G=c=1$, and signature $(-,+,+,+)$.
The action is
\begin{equation}
\begin{aligned}
S[g_{\mu\nu},\Psi] &= \frac{1}{16\pi}\int \dd^4x\,\sqrt{-g}\,\bigl(R+\lambda T^2\bigr)\\
&\quad+ \int \dd^4x\,\sqrt{-g}\,\Lm(g_{\mu\nu},\Psi),
\end{aligned}
\label{eq:action}
\end{equation}
where $\Psi$ denotes the matter fields and
\begin{equation}
\begin{aligned}
T_{\mu\nu} &\equiv -\frac{2}{\sqrt{-g}}\frac{\delta(\sqrt{-g}\,\Lm)}{\delta g^{\mu\nu}},\\
T &\equiv g^{\mu\nu}T_{\mu\nu}.
\end{aligned}
\label{eq:stressdef}
\end{equation}
For the model \eqref{eq:intro_model} one has
\begin{equation}
 f_R=1,
 \qquad
 f_T=2\lambda T,
\label{eq:fR_fT}
\end{equation}
so the metric field equations contain no derivatives higher than second order.

For matter Lagrangians that depend on the metric but not its derivatives, the variation of \eqref{eq:action} gives \cite{Harko2011}
\begin{equation}
G_{\mu\nu}
=
8\pi T_{\mu\nu}-2\lambda T\bigl(T_{\mu\nu}+\Theta_{\mu\nu}\bigr)
+\frac{1}{2}\lambda T^2 g_{\mu\nu},
\label{eq:fieldeq}
\end{equation}
where
\begin{equation}
\Theta_{\mu\nu}
\equiv
g^{\alpha\beta}\frac{\delta T_{\alpha\beta}}{\delta g^{\mu\nu}}.
\label{eq:Thetadef}
\end{equation}
In general, $\Theta_{\mu\nu}$ also contains the metric-variation term
\[
-2g^{\alpha\beta}\frac{\partial^2\Lm}
{\partial g^{\mu\nu}\partial g^{\alpha\beta}}.
\]
For the phenomenological algebraic closures studied here
we \emph{assume}
\begin{equation}
\Theta_{\mu\nu}=-2T_{\mu\nu}+\Lm g_{\mu\nu},
\label{eq:Thetafluid}
\end{equation}
as the defining closure relation. Substituting $\Lm=\Pbar$ or $p_r$ is not by itself a microscopic variational derivation of this assumption. Subject to it, Eq.~\eqref{eq:fieldeq} takes the Einstein-like form
\begin{equation}
G_{\mu\nu}=\mathcal{A}(T)T_{\mu\nu}+\mathcal{B}(T,\Lm)g_{\mu\nu},
\label{eq:ABeq}
\end{equation}
with
\begin{equation}
\mathcal{A}(T)=8\pi+2\lambda T,
\qquad
\mathcal{B}(T,\Lm)=\frac12\lambda T^2-2\lambda T\Lm.
\label{eq:ABdef}
\end{equation}

We model the matter source as an anisotropic fluid,
\begin{equation}
T^{\mu}{}_{\nu}=\mathrm{diag}(-\rho,p_r,p_t,p_t),
\qquad
T=-\rho+p_r+2p_t,
\label{eq:anisotropic}
\end{equation}
with either
\begin{equation}
\Lm=\Pbar\equiv \frac{p_r+2p_t}{3},
\qquad \text{or} \qquad
\Lm=p_r.
\label{eq:Lmchoices}
\end{equation}
The first choice is algebraically symmetric in the angular sector, while the second is often used in the anisotropic-fluid literature. Since both lead to inequivalent trace reconstructions, we keep them separate throughout.

Defining the effective tensor through
\begin{equation}
\begin{aligned}
G_{\mu\nu} &= 8\pi T^{\mathrm{eff}}_{\mu\nu},\\
T^{\mathrm{eff}\,\mu}{}_{\nu} &= \mathrm{diag}(-\rho_{\rm eff},p_{r,{\rm eff}},p_{t,{\rm eff}},p_{t,{\rm eff}}).
\end{aligned}
\label{eq:Teff}
\end{equation}
we obtain
\begin{subequations}
\label{eq:effvars}
\begin{align}
8\pi\rho_{\rm eff}&=\mathcal{A}(T)\rho-\mathcal{B}(T,\Lm),
\\
8\pi p_{r,{\rm eff}}&=\mathcal{A}(T)p_r+\mathcal{B}(T,\Lm),
\\
8\pi p_{t,{\rm eff}}&=\mathcal{A}(T)p_t+\mathcal{B}(T,\Lm).
\end{align}
\end{subequations}
Introducing the dimensionless trace variable
\begin{equation}
\chi(r)\equiv \frac{\lambda T(r)}{4\pi},
\qquad
\mathcal{A}(T)=8\pi(1+\chi),
\label{eq:chi}
\end{equation}
we obtain the exact relations
\begin{subequations}
\label{eq:signtransfer}
\begin{align}
\rho_{\rm eff}+p_{r,{\rm eff}}&=(1+\chi)(\rho+p_r),
\\
\rho_{\rm eff}+p_{t,{\rm eff}}&=(1+\chi)(\rho+p_t),
\\
p_{r,{\rm eff}}-p_{t,{\rm eff}}&=(1+\chi)(p_r-p_t).
\end{align}
\end{subequations}
These identities show immediately that the singular inversion surface is
\begin{equation}
1+\chi=0.
\label{eq:critical}
\end{equation}

The static, spherically symmetric wormhole metric is written in Morris--Thorne form,
\begin{equation}
\dd s^2=-\ee^{2\Phi(r)}\dd t^2+\frac{\dd r^2}{1-b(r)/r}+r^2\dd\Omega^2,
\label{eq:metric}
\end{equation}
with throat conditions
\begin{equation}
\begin{aligned}
b(r_0)&=r_0,\qquad b'(r_0)<1,\qquad b(r)<r\ (r>r_0),\\
\Phi(r)&\text{ finite}.
\end{aligned}
\label{eq:throat}
\end{equation}
The effective source generated by \eqref{eq:metric} is
\begin{subequations}
\label{eq:effsource}
\begin{align}
8\pi\rho_{\rm eff} &= \frac{b'}{r^2},\\
8\pi p_{r,{\rm eff}} &= -\frac{b}{r^3}+\frac{2}{r}\left(1-\frac{b}{r}\right)\Phi',\\
8\pi p_{t,{\rm eff}} &= \left(1-\frac{b}{r}\right)\left(\Phi''+\Phi'^2+\frac{\Phi'}{r}\right)\\
&\quad+\frac{b-rb'}{2r^2}\left(\Phi'+\frac{1}{r}\right).
\end{align}
\end{subequations}
These equations define the geometry-to-source map that underlies the whole construction.

Taking the covariant divergence of Eq.~\eqref{eq:ABeq} gives the complete
nonconservation law for this closure system,
\begin{equation}
\begin{aligned}
\nabla^\mu T_{\mu\nu}
&=\frac{\lambda}{4\pi+\lambda T}
\Big[T\nabla_\nu\Lm+\left(\Lm-\frac{T}{2}\right)\nabla_\nu T\\
&\hspace{34mm}-T_{\mu\nu}\nabla^\mu T\Big].
\end{aligned}
\label{eq:divlaw}
\end{equation}
It follows directly from $\nabla^\mu G_{\mu\nu}=0$, without dividing by $T$;
it is therefore also well defined where the trace itself vanishes, provided
$1+\chi\ne0$. The trace-gradient terms are essential for a nonlinear trace
coupling \cite{BarrientosRubilar2014}.
For the static diagonal source, the radial balance is
\begin{equation}
p_r'+\Phi'(\rho+p_r)-\frac{2}{r}(p_t-p_r)=\mathcal F_T,
\label{eq:modifiedTOV}
\end{equation}
where
\begin{equation}
\mathcal F_T=\frac{\lambda}{4\pi+\lambda T}
\left[T\Lm'+\left(\Lm-p_r-\frac{T}{2}\right)T'\right].
\label{eq:force}
\end{equation}
The additional term proportional to $(\Lm-p_r)T'$ vanishes for $\Lm=p_r$
but not generally for $\Lm=\Pbar$. We use Eq.~\eqref{eq:modifiedTOV} as a
consistency check of the algebraic reconstruction, not as an independent
evolution equation.

For later use it is convenient to record the effective throat data implied by Eqs.~\eqref{eq:effsource} and \eqref{eq:throat}. At $r=r_0$ one finds
\begin{equation}
\begin{aligned}
8\pi\rho_{\rm eff}(r_0) &= \frac{b'(r_0)}{r_0^2},\\
8\pi p_{r,{\rm eff}}(r_0) &= -\frac{1}{r_0^2},\\
8\pi\bigl[\rho_{\rm eff}+p_{r,{\rm eff}}\bigr]_{r_0} &= \frac{b'(r_0)-1}{r_0^2}<0.
\end{aligned}
\label{eq:throat_effective_data}
\end{equation}
so the effective radial NEC is always violated at the throat. Moreover, when $0<b'(r_0)<1$ the effective throat density is positive and the effective radial equation-of-state parameter is
\begin{equation}
\omega^{\rm eff}_{r,0}=\frac{p_{r,{\rm eff}}(r_0)}{\rho_{\rm eff}(r_0)}=-\frac{1}{b'(r_0)}<-1.
\label{eq:omega_eff_throat}
\end{equation}
These statements concern the effective throat source. Positivity of the physical throat density must be assessed separately after reconstruction.

\section{Compact-Support Schwarzschild Deformation}
\label{sec:compact}
Choose $0<2M<r_0<\Rc$, and introduce
\begin{equation}
\Delta=\Rc-r_0,\qquad x=\frac{r-r_0}{\Delta}.
\label{eq:xdef}
\end{equation}
The core corresponds to $0\le x\le1$. For positive integers $m,n$, define
\begin{equation}
\sigma(x)=\begin{cases}
\exp\!\left(1-\dfrac{1}{1-x}\right),&0\le x<1,\\
0,&x\ge1,
\end{cases}
\label{eq:sigma}
\end{equation}
and
\begin{equation}
q_n=x^n\sigma,\qquad
\mathcal I_n(x)=\int_0^x q_n(y)\dd y,\qquad
\mathcal J_n(x)=\int_x^1 q_n(y)\dd y.
\label{eq:qIJ}
\end{equation}
The functions $q_n$ and $\mathcal J_n$ vanish to all orders at $x=1$.
The positive-order derivatives of $\mathcal I_n$ vanish there, whereas
$I_n\equiv\mathcal I_n(1)>0$. These are smooth functions on the radial
half-line, with one-sided throat derivatives.

For $0<s_0<1$, take
\begin{equation}
b(r)=\begin{cases}
r_0+\Delta[s_0\mathcal I_0(x)-\beta\mathcal I_m(x)],&r_0\le r\le\Rc,\\
2M,&r\ge\Rc,
\end{cases}
\label{eq:shape}
\end{equation}
with
\begin{equation}
\beta=\frac{s_0I_0+(r_0-2M)/\Delta}{I_m},\qquad
b'(r)=\sigma(x)(s_0-\beta x^m).
\label{eq:beta}
\end{equation}
Then $b(r_0)=r_0$, $b'(r_0)=s_0<1$, and $b(\Rc)=2M$.
Because $b'\le s_0<1$, the gap $r-b(r)$ grows strictly from zero, proving
$b(r)<r$ for $r>r_0$. Also $\beta>s_0$, and $b'$ changes sign once at
$x_t=(s_0/\beta)^{1/m}$; hence $b(r)\ge2M>0$.

Write $\nu=-\varepsilon$, and set
\begin{equation}
\Phi(r)=\frac12\ln\!\left(1-\frac{2M}{r}\right)+\delta\Phi(r),\qquad
\delta\Phi(r)=\begin{cases}
\nu\mathcal J_n(x),&r_0\le r\le\Rc,\\
0,&r\ge\Rc.
\end{cases}
\label{eq:Phi_core}
\end{equation}
The lapse is positive because $r_0>2M$ and $\mathcal J_n$ is bounded.
With this sign convention the core derivatives are
\begin{equation}
\begin{aligned}
\Phi'&=\frac{M}{r(r-2M)}-\frac{\nu}{\Delta}q_n(x),\\
\Phi''&=-\frac{M(2r-2M)}{r^2(r-2M)^2}-\frac{\nu}{\Delta^2}q_n'(x).
\end{aligned}
\label{eq:Phi_derivatives}
\end{equation}
Equivalently, both compact contributions have plus signs when expressed
using $\varepsilon=-\nu$.

Let $\delta b=b-2M$. The boundary condition in Eq.~\eqref{eq:beta} and
the flat cutoff imply
\begin{equation}
\delta b^{(k)}(\Rc^-)=\delta\Phi^{(k)}(\Rc^-)=0,
\qquad k=0,1,2,\ldots.
\label{eq:flatmatching}
\end{equation}
Thus the metric agrees to all orders with the Schwarzschild exterior, and
the second-order field equations contain no surface distribution at $\Rc$.

Near the throat,
\begin{equation}
\begin{aligned}
1-b/r&=\frac{1-s_0}{r_0}(r-r_0)+O((r-r_0)^2),\\
\ell(r)&=\pm2\sqrt{\frac{r_0}{1-s_0}}\sqrt{r-r_0}
+O((r-r_0)^{3/2}),
\end{aligned}
\label{eq:throatexpansion}
\end{equation}
where $\ell=\pm\int_{r_0}^r[1-b(u)/u]^{-1/2}\dd u$.
The proper-distance coordinate gives a smooth two-ended extension, and
the embedding flare-out is $(\dd^2r/\dd z^2)_{r_0}=(1-s_0)/(2r_0)>0$.

\begin{figure}[tbp]
\centering\includegraphics[width=\textwidth]{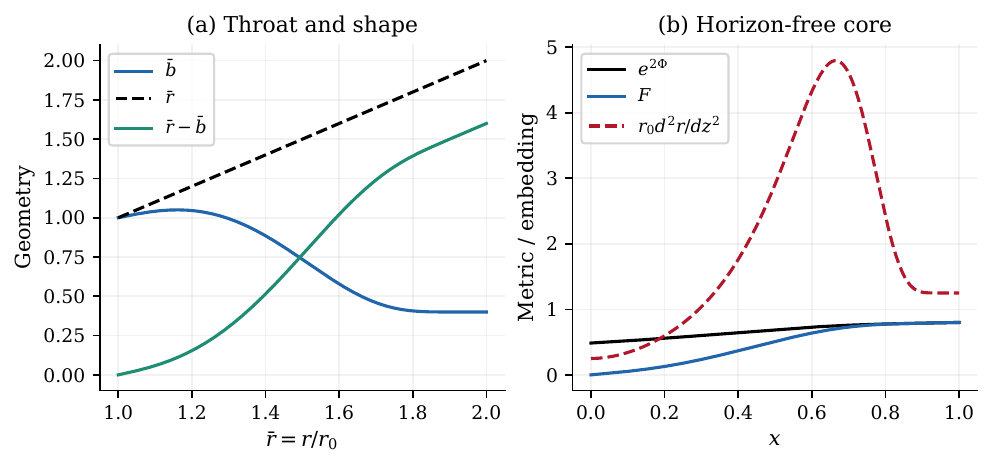}
\caption{Geometry for $(\mu,\Delta_c,s_0,\nu;m,n)=(0.4,1,0.5,-1;2,1)$.
(a) The shape function stays below the areal radius away from the throat.
(b) The lapse remains positive and the throat flares out. Dimensionless
parameters are defined in Eq.~\eqref{eq:dimensionlesspars}.}
\label{fig:geometry_and_throat}
\end{figure}
\begin{figure}[tbp]
\centering\includegraphics[width=\textwidth]{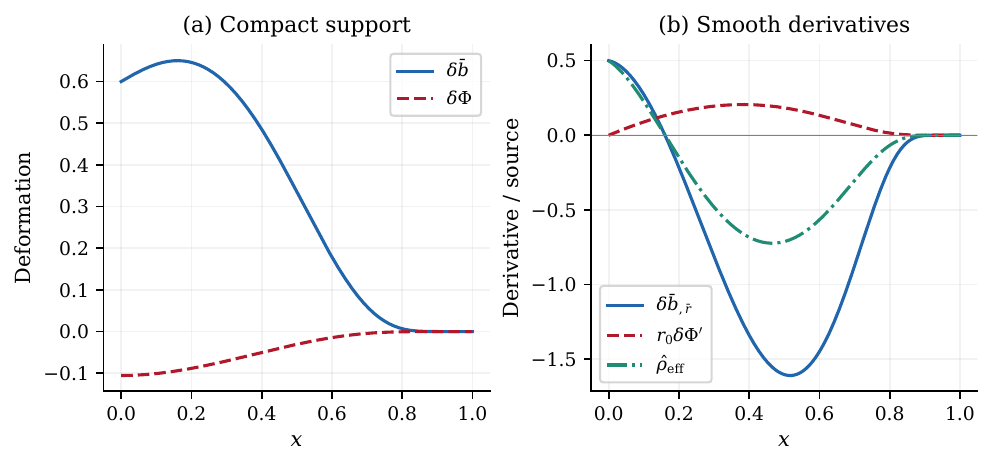}
\caption{Compact-support matching for the same geometry.
(a) Both deformations vanish at $x=1$.
(b) The redshift deformation has a positive radial derivative for $\nu=-1$,
consistent with Eq.~\eqref{eq:Phi_derivatives}; the displayed derivatives and
effective density vanish at the compact boundary.}
\label{fig:compact_support_matching}
\end{figure}

\section{Matter Reconstruction and Branch Continuation}
\label{sec:reconstruction}
Once the geometry is fixed, the physical anisotropic variables are reconstructed locally from the effective source. Solving Eqs.~\eqref{eq:effvars} together with the trace definition gives the branch-independent inversion formulas
\begin{subequations}
\label{eq:unifiedinversion}
\begin{align}
\rho&=
\frac{3\rho_{\rm eff}+p_{r,{\rm eff}}+2p_{t,{\rm eff}}}{4(1+\chi)}
-\frac{\pi}{\lambda}\chi,
\\
p_r&=
\frac{\rho_{\rm eff}+3p_{r,{\rm eff}}-2p_{t,{\rm eff}}}{4(1+\chi)}
+\frac{\pi}{\lambda}\chi,
\\
p_t&=
\frac{\rho_{\rm eff}-p_{r,{\rm eff}}+2p_{t,{\rm eff}}}{4(1+\chi)}
+\frac{\pi}{\lambda}\chi.
\end{align}
\end{subequations}
Thus the entire matter reconstruction problem reduces to determining $\chi(r)$.

For $\Lm=\Pbar$, combining the trace equation with the effective source gives the cubic
\begin{equation}
\chi^3+2\chi^2+
\left(1-\frac{\lambda}{\pi}\bar p_{\rm eff}\right)\chi
-\frac{\lambda}{4\pi}T_{\rm eff}=0,
\label{eq:cubicP}
\end{equation}
where
\begin{equation}
\bar p_{\rm eff}\equiv\frac{p_{r,{\rm eff}}+2p_{t,{\rm eff}}}{3},
\qquad
T_{\rm eff}\equiv -\rho_{\rm eff}+p_{r,{\rm eff}}+2p_{t,{\rm eff}}.
\label{eq:Teffbarp}
\end{equation}
For $\Lm=p_r$, one obtains instead
\begin{equation}
\chi^3+2\chi^2+
\left(1-\frac{\lambda}{\pi}p_{r,{\rm eff}}\right)\chi
-\frac{\lambda}{4\pi}T_{\rm eff}=0.
\label{eq:cubicpr}
\end{equation}

For a unified description, define
\begin{equation}
\xi_s=\frac{\lambda}{\pi}\Pi^{(s)}_{\rm eff},\quad
\eta=\frac{\lambda}{4\pi}T_{\rm eff},\quad
\Pi^{(\Pbar)}_{\rm eff}=\bar p_{\rm eff},\quad
\Pi^{(r)}_{\rm eff}=p_{r,{\rm eff}}.
\label{eq:xi_eta_def}
\end{equation}
The cubic and its root Jacobian are
\begin{align}
P_s(\chi,r)&=\chi^3+2\chi^2+(1-\xi_s)\chi-\eta=0,
\label{eq:generic_cubic_main}\\
J_s(\chi,r)&\equiv\partial_\chi P_s=3\chi^2+4\chi+1-\xi_s.
\label{eq:jacobian}
\end{align}
Its discriminant can be evaluated without cancellation of vacuum terms as
\begin{equation}
\mathfrak D_s=4\xi_s(1-\xi_s)^2+(36\xi_s-4)\eta-27\eta^2.
\label{eq:discriminant_main}
\end{equation}
Positive and negative discriminants denote three and one distinct real roots,
respectively. A zero discriminant need not concern the selected root.

At $r=\Rc$ the cubic is $\chi(\chi+1)^2=0$.
The vacuum root $\chi=0$ is simple, with $J_s=1$, so the implicit function
theorem gives a unique smooth local continuation. The double root $\chi=-1$
has a vanishing inversion denominator and is excluded from this vacuum branch.
Away from the boundary, a simple selected root obeys
\begin{equation}
\chi_s'=\frac{\xi_s'\chi_s+\eta'}{J_s(\chi_s,r)}.
\label{eq:branch_derivative}
\end{equation}
We follow this continuation only while it stays real, simple, and satisfies
$1+\chi_s>0$. A selected-root fold, $P_s=J_s=0$, is distinct from the inversion
surface $1+\chi_s=0$. A disconnected real root is not substituted after a fold.
Special nonsimple continuations, if present, require a separate regularity
analysis and are not included in the strict admissible set used here.

For a full continuation on the closed core define
\begin{equation}
\Xi_s=\min_{[r_0,\Rc]}(1+\chi_s),\qquad
H_s=\min_{[r_0,\Rc]}J_s(\chi_s,r),\qquad
\mathcal R_s=\rho^{(s)}(r_0).
\label{eq:diagnostics}
\end{equation}
The strict regular set has $\Xi_s>0$ and $H_s>0$.
These global minima and the physical throat values are not assigned to a
continuation that terminates before reaching the throat. We retain the
following phase notation: I denotes a full regular branch with
$\mathcal R_s>0$; II denotes a full regular branch with $\mathcal R_s\le0$;
III denotes a critical boundary where a selected-root Jacobian or inversion
margin vanishes; and IV denotes parameters without a full regular simple
continuation under these criteria. Phase III is not a region of finite width.

A useful explicit example is the baseline throat at $\mu=0.4$, $s_0=0.5$,
$\nu=0$, for which the $\Pbar$ cubic becomes
\begin{equation}
\chi^3+2\chi^2+\left(1+\frac{\alpha}{9}\right)\chi
+\frac{5\alpha}{24}=0.
\label{eq:baseline_throat_cubic}
\end{equation}
Its discriminant is
$-\alpha(256\alpha^2+20403\alpha-18144)/46656$.
The selected positive-coupling fold occurs at
\begin{equation}
\alpha_f\simeq\AlphaCrit,\qquad
\chi_f\simeq\ChiCrit,\qquad
1+\chi_f\simeq\UCrit.
\label{eq:explicit_fold}
\end{equation}
Thus $H_s$ can vanish with $\Xi_s$ bounded away from zero. At $\alpha=0$,
the vanishing discriminant instead refers to the unselected vacuum double
root; the selected GR branch is regular.

The formal zero-throat-density candidates, for the families with $n\ge1$, are
\begin{align}
\alpha_{\Pbar}^{(\rho)}&=\frac{48s_0}{[10s_0-(1-s_0)\varphi_0]^2},
\label{eq:rho_zero_P}\\
\alpha_r^{(\rho)}&=-\frac{16s_0[4-2s_0+(1-s_0)\varphi_0]}
{[2s_0+(1-s_0)\varphi_0][4+2s_0+(1-s_0)\varphi_0]^2},
\label{eq:rho_zero_r}
\end{align}
where $\varphi_0=\mu/[2(1-\mu)]$ and $\alpha$ is defined below.
They represent physical transitions only when they lie on the selected full
regular branch. For example, the baseline $\mu=0.4$, $s_0=0.5$ gives
$\alpha_{\Pbar}^{(\rho)}\simeq1.02735$, beyond the fold in
Eq.~\eqref{eq:explicit_fold}. It is therefore not an admissible phase-I/II
transition on that slice.

The sign-transfer identities in Eq.~\eqref{eq:signtransfer} give
\begin{equation}
(\rho+p_r)_{r_0}=\frac{s_0-1}{8\pi r_0^2[1+\chi_s(r_0)]}<0
\label{eq:throatNEC}
\end{equation}
on every full regular branch. Positive \emph{throat} density then implies a
radial phantom equation of state there; it does not imply positive density
throughout the core.

For $\nu=0$ the effective radial NEC has the cancellation-free form
\begin{equation}
8\pi(\rho_{\rm eff}+p_{r,{\rm eff}})
=\frac1{r^2}\left[b'-\frac{b-2M}{r-2M}\right].
\label{eq:stable_nec}
\end{equation}
It is strictly negative for $r_0\le r<\Rc$.
To see this, put $N=(r-2M)b'-(b-2M)$. Before the single maximum of $b$,
$b''<0$, so $N'=(r-2M)b''<0$ and
$N(r_0)=(r_0-2M)(s_0-1)<0$. After that maximum, $b'\le0$ and $b-2M>0$,
which also gives $N<0$. The NEC reaches zero only at the compact boundary.
On a regular matter branch it has the same physical sign.
For $\nu\ne0$, Eq.~\eqref{eq:stable_nec} acquires the term
$2(1-b/r)\delta\Phi'/r$. The profiles are evaluated with this term retained;
no apparent floating-point tail zero is interpreted as a physical endpoint.

\section{Numerical Results}
\label{sec:numerics}
We use dimensionless variables
\begin{equation}
\begin{aligned}
\bar r&=r/r_0,\qquad \bar b=b/r_0,\qquad
\mu=2M/r_0,\\
\Delta_c&=(\Rc-r_0)/r_0,\qquad \nu=-\varepsilon,\qquad
\alpha=\lambda/(8\pi^2r_0^2),
\end{aligned}
\label{eq:dimensionlesspars}
\end{equation}
and $\hat X=8\pi r_0^2X$ for every physical or effective density or pressure.
Thus the cubic coefficients become $\xi_s=\alpha\hat\Pi^{(s)}_{\rm eff}$
and $\eta=\alpha\hat T_{\rm eff}/4$. At $\alpha=0$ we set
$\chi=0$ and $(\rho,p_r,p_t)=(\rho_{\rm eff},p_{r,{\rm eff}},p_{t,{\rm eff}})$
directly, avoiding the removable $1/\lambda$ expressions in
Eq.~\eqref{eq:unifiedinversion}.

The baseline family fixes $(\Delta_c,s_0,\nu)=(1,0.5,0)$ with $(m,n)=(2,1)$.
Family B fixes $(\mu,\nu)=(0.9,-1)$, also with $(m,n)=(2,1)$, and varies the throat slope
or core width. The numerical branch starts at the vacuum boundary, and
discriminant events are resolved without exchanging a terminating root
for a different real solution. Full global margins are reported only when
the selected continuation reaches the throat.

\begin{figure}[tbp]
\centering\includegraphics[width=\textwidth]{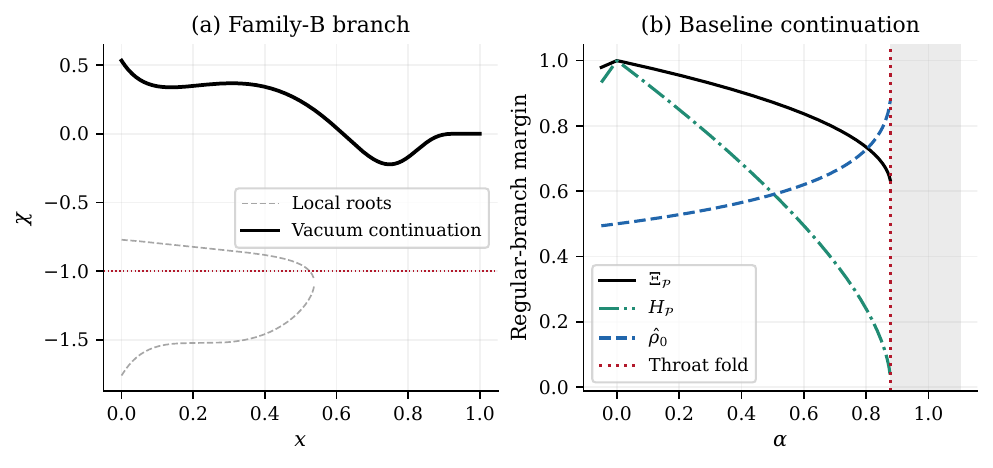}
\caption{(a) Local real roots and selected continuation for Family B at
$(\alpha,\mu,\Delta_c,s_0,\nu)=(0.75,0.9,0.5,0.1,-1)$.
Separate root segments are not joined across changes in multiplicity.
(b) Baseline $\mu=0.4$ scan: the Jacobian margin $H_{\Pbar}$ approaches zero
at the throat fold while $\Xi_{\Pbar}$ remains positive.
Global diagnostics are not extended into the shaded inadmissible region.}
\label{fig:branch_structure_and_scan}
\end{figure}

\begin{figure}[tbp]
\centering\includegraphics[width=\textwidth]{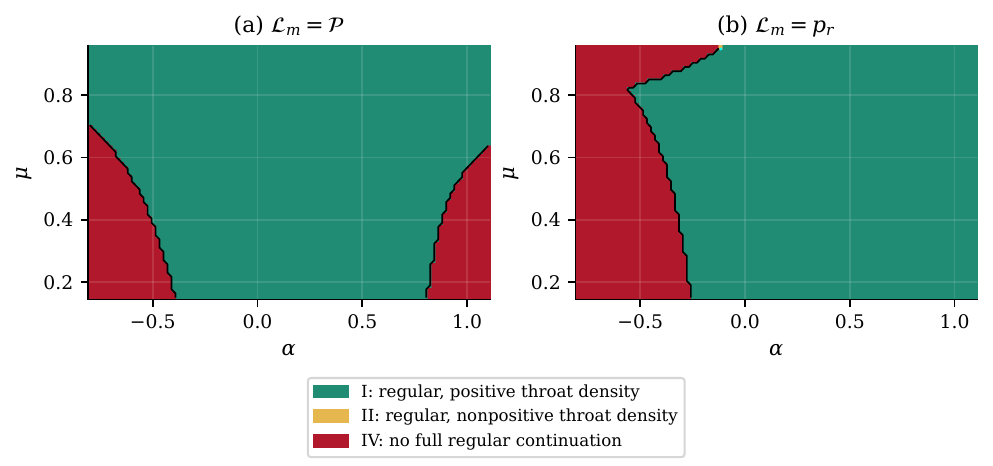}
\caption{Baseline phase maps for $(\Delta_c,s_0,\nu;m,n)=(1,0.5,0;2,1)$.
The grid has 101 coupling values on $[-0.8,1.1]$, 61 compactness values on
$[0.15,0.95]$, and 1601 core points per reconstruction. Black contours
approximate the boundary between full regular continuations and phase IV;
they are limited by the parameter-grid resolution. Phase III refers to
critical boundary points, not a colored band of finite width.}
\label{fig:baseline_phase_maps}
\end{figure}

Figure~\ref{fig:baseline_phase_maps} displays the baseline parameter slices.
The two prescriptions give different admissibility domains. For example,
at $\mu=0.4$ the $p_r$ prescription has a full regular branch at $\alpha=1$,
whereas the $\Pbar$ continuation has already encountered a fold.
No phase-II points occur in the displayed baseline $\Pbar$ grid.
The $p_r$ grid contains a small negative-coupling, high-compactness phase-II
sector; its small size should not be interpreted beyond the sampled resolution.
The analytic density candidates in Eqs.~\eqref{eq:rho_zero_P} and
\eqref{eq:rho_zero_r} must therefore be checked against the continuation.

\begin{figure}[tbp]
\centering\includegraphics[width=\textwidth]{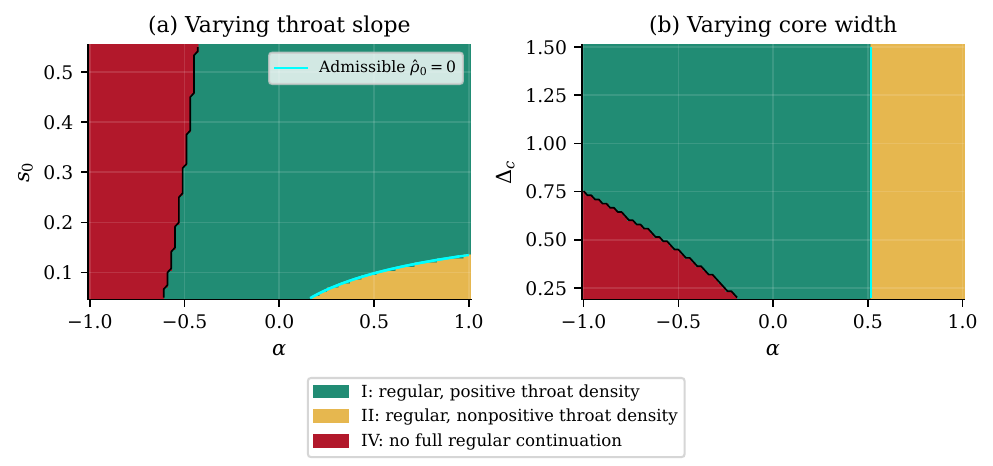}
\caption{Family-B $\Pbar$ phase maps for $\mu=0.9$, $\nu=-1$, $m=2$, $n=1$.
(a) The slope slice fixes $\Delta_c=0.5$ and uses $0.05\le s_0\le0.55$.
(b) The width slice fixes $s_0=0.1$ and uses $0.2\le\Delta_c\le1.5$.
Each panel uses 101 coupling values on $[-1,1]$, 61 ordinate values, and
1601 radial points. The cyan curve in (a) is the admissible portion of
Eq.~\eqref{eq:rho_zero_P}; black boundaries have the parameter-grid accuracy.}
\label{fig:familyB_phase_maps}
\end{figure}

Family B provides regular negative-throat-density examples with the redshift
derivatives specified in Eq.~\eqref{eq:Phi_derivatives}.
Its phase maps are shown in Fig.~\ref{fig:familyB_phase_maps}.
The cyan density transition in the slope slice is restricted to the portion
of the analytic candidate that belongs to a full regular branch.
The example at $\alpha=0.75$, $s_0=0.1$, $\Delta_c=0.5$ has
$\Xi_{\Pbar}\simeq\XiB$, $H_{\Pbar}\simeq\HB$, and
$\hat\rho_0\simeq\RhoB$. It remains admissible even though the throat
density is negative; positive throat density is an additional preference,
not a necessary condition for geometric existence.

\begin{figure}[tbp]
\centering\includegraphics[width=\textwidth]{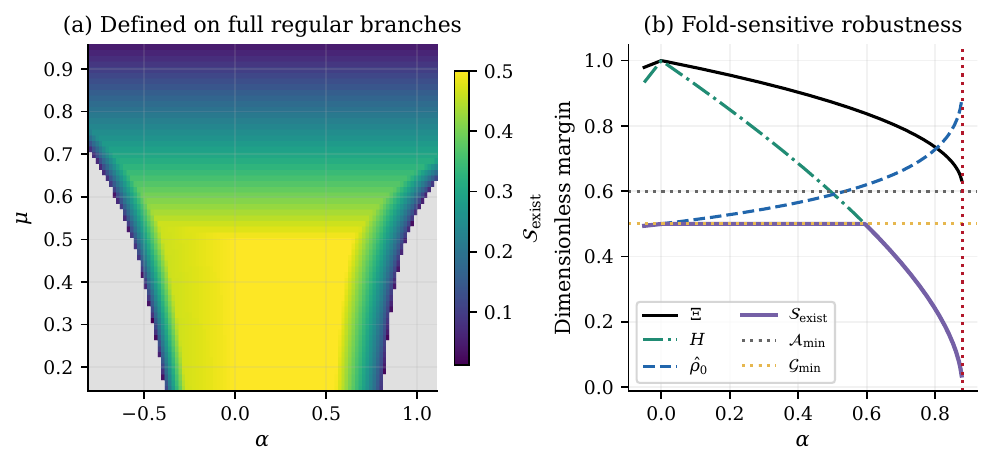}
\caption{(a) The baseline $\Pbar$ robustness index from Eq.~\eqref{eq:Sexist_main}.
Gray indicates that a full regular continuation was not obtained, rather than
a continuation of the index to disconnected roots.
(b) Baseline $\mu=0.4$ margins. The Jacobian margin approaches zero at the
fold, while the geometric margins remain positive.}
\label{fig:existence_robustness}
\end{figure}

For a fully defined, dimensionless robustness diagnostic we set
\begin{equation}
\begin{aligned}
\mathcal A_{\min}&=\min_{0\le x\le1}\ee^{2\Phi(x)},\\
\mathcal G_{\min}&=\min_{0\le x\le1}g(x),\qquad
g(x)=\frac{\bar r-\bar b}{\bar r-1},\quad g(0)=1-s_0,\\
\mathcal S_{\rm exist}&=\min\{\Xi_s,H_s,\mathcal A_{\min},
\mathcal G_{\min},\hat\rho_0\}.
\end{aligned}
\label{eq:Sexist_main}
\end{equation}
The normalized gap removes the obligatory zero of $r-b$ at the throat.
The inclusion of $H_s$ makes a selected-root fold visible even when $\Xi_s$
is positive. Figure~\ref{fig:existence_robustness} defines this index only
on full regular branches; gray regions have no assigned global index.
A negative value on a regular phase-II branch only reflects the throat-density
preference. This conventional minimum compares dimensionless margins within
the displayed families; it is not an invariant measure of dynamical stability.

Table~\ref{tab:benchmark_points} records the three full regular benchmarks.
All use $(m,n)=(2,1)$. We report the geometric proper core radius
$\ell_c=\int_{r_0}^{\Rc}(1-b/r)^{-1/2}\dd r$, not a zero inferred from the
numerically vanishing outer NEC tail. For the baseline, the proof following
Eq.~\eqref{eq:stable_nec} establishes that the radial NEC-violating interval
reaches the compact boundary. Its one-sided proper extent is
$\ell_c/r_0\simeq\LBase$; the reflected core has twice that extent.
The Family-B core radius is $\ell_c/r_0\simeq\LFamilyB$.
Its displayed radial NEC profile is also negative across the sampled open core,
but the analytic all-core sign proof given above is specific to $\nu=0$.

\begin{table}[tbp]
\caption{Full regular benchmarks. Tuples list
$(\alpha,\mu,\Delta_c,s_0,\nu)$; hatted stresses have the normalization stated
above. All entries use $(m,n)=(2,1)$.}
\label{tab:benchmark_points}
\centering\fontsize{8}{10}\selectfont\setlength{\tabcolsep}{3pt}
\begin{tabular}{@{}clrrrr@{}}
\toprule
Phase / closure & Tuple & $\Xi_s$ & $H_s$ & $\hat\rho_0$ & $\chi_0$\\
\midrule
I / $\Pbar$ & $(.5,.4,1,.5,0)$ & 0.872461 & 0.594198 & 0.589381 & -0.127539\\
I / $p_r$ & $(1,.4,1,.5,0)$ & 0.882930 & 1.000000 & 0.447409 & -0.117070\\
II / $\Pbar$ & $(.75,.9,.5,.1,-1)$ & 0.776842 & 0.541603 & -0.014963 & 0.531534\\
\bottomrule
\end{tabular}

\end{table}

\begin{figure}[tbp]
\centering\includegraphics[width=\textwidth]{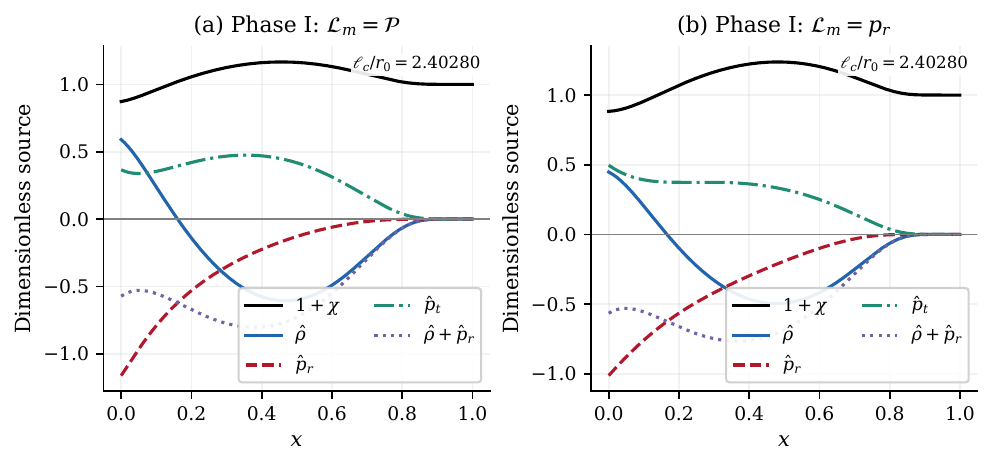}
\caption{Phase-I physical source profiles for the two baseline benchmarks
in Table~\ref{tab:benchmark_points}. Both have positive throat density;
the density is not required to remain positive throughout the core.
The radial NEC is negative throughout $0\le x<1$, with exact zero at $x=1$.
The annotation gives the proper core radius, not a floating-point NEC crossing.}
\label{fig:phaseI_profiles}
\end{figure}

\begin{figure}[tbp]
\centering\includegraphics[width=\textwidth]{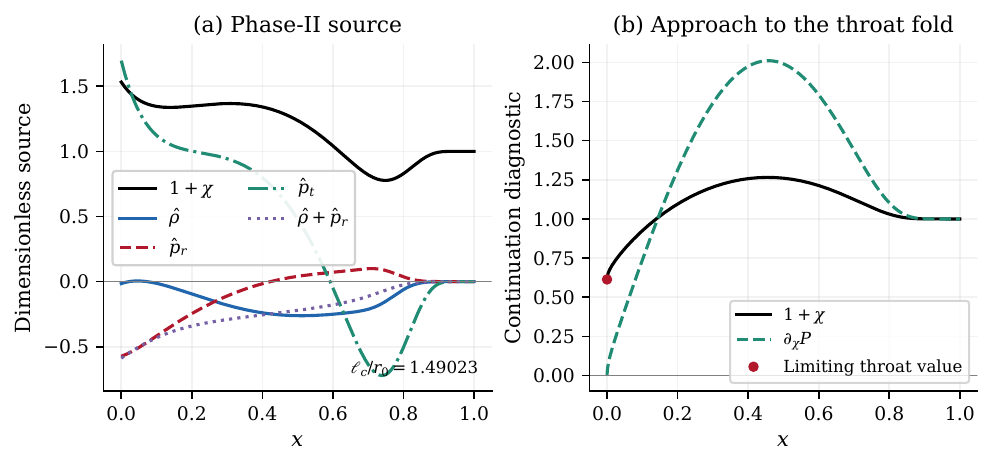}
\caption{(a) Family-B phase-II profiles for the third row of
Table~\ref{tab:benchmark_points}.
(b) A baseline $\Pbar$ branch at $\alpha=\alpha_f-10^{-6}$ approaches the
throat fold. The Jacobian tends to zero at the limiting throat point, whereas
$1+\chi$ tends to approximately $0.613573$; the two degeneracy criteria are distinct.}
\label{fig:phaseII_and_critical_profiles}
\end{figure}

The two phase-IV examples illustrate branch termination, not a geometric
singularity. At $\alpha=1$ for $\Lm=\Pbar$, the selected pair coalesces at
$x_f\simeq\FoldXP$, with $1+\chi_f\simeq\FoldUP$.
At $\alpha=-0.5$ for $\Lm=p_r$, the fold lies at
$x_f\simeq\FoldXR$, with $1+\chi_f\simeq\FoldUR$.
In both cases the strict physical continuation is plotted only on
$x_f<x\le1$, including its finite limiting value at the fold.
No physical throat density, full-core $\Xi_s$, or full-core $H_s$ is assigned
to these examples. The smooth effective geometry remains defined everywhere.

\begin{table}[tbp]
\caption{Fold-limited baseline examples, with $(\mu,\Delta_c,s_0,\nu)=(.4,1,.5,0)$.
Global matter diagnostics are undefined because the selected simple branch
does not reach the throat.}
\label{tab:fold_points}
\centering\fontsize{8}{10}\selectfont
\begin{tabular}{@{}ccccc@{}}\toprule
Closure & $\alpha$ & $x_f$ & $1+\chi_f$ & Termination\\\midrule
$\Pbar$ & $1$ & $\FoldXP$ & $\FoldUP$ & Selected-root fold\\
$p_r$ & $-.5$ & $\FoldXR$ & $\FoldUR$ & Selected-root fold\\\bottomrule
\end{tabular}
\end{table}
\begin{figure}[tbp]
\centering\includegraphics[width=\textwidth]{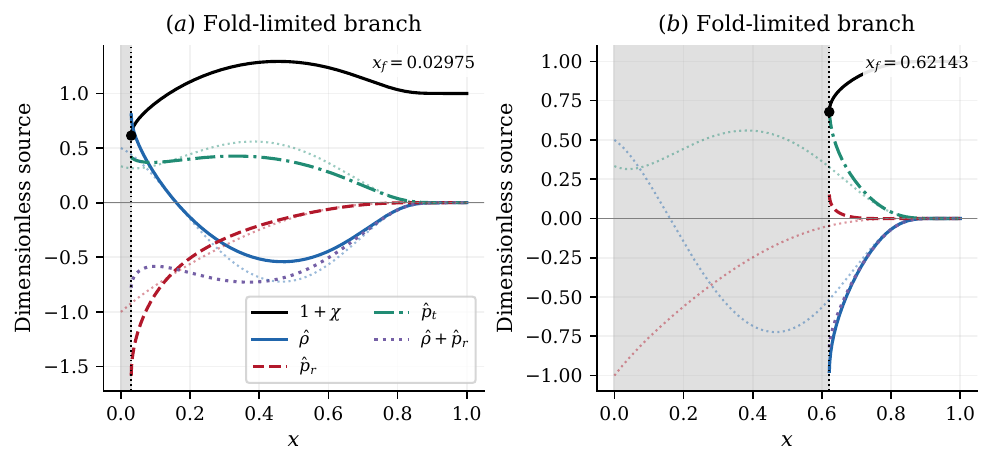}
\caption{Fold-limited continuations for Table~\ref{tab:fold_points}.
(a) $\Lm=\Pbar$, $\alpha=1$; (b) $\Lm=p_r$, $\alpha=-0.5$.
The selected root stops at the marked fold. Shaded inner regions have no
assigned physical continuation; faint dotted curves show effective sources
only. No line joins the selected root to a disconnected real root.}
\label{fig:phaseIV_truncated_profiles}
\end{figure}
\section{Regularity, Asymptotics, and Physical Interpretation}
\label{sec:regularity}

The compact-support construction makes the smoothness of the full wormhole manifold transparent. Since $b(r)$ and $\Phi(r)$ are $C^\infty$ on $[r_0,\infty)$ and satisfy the throat conditions \eqref{eq:throat}, the proper-distance coordinate extends smoothly through $r=r_0$, producing a smooth two-ended geometry. The compact-support boundary $r=\Rc$ is not a matching hypersurface in the shell sense; instead, the metric and all its derivatives agree with the exact Schwarzschild exterior there.

Invariant regularity follows directly from the effective Einstein system. The Ricci scalar and quadratic Ricci invariant are
\begin{equation}
\begin{aligned}
R &= 8\pi\bigl(\rho_{\rm eff}-p_{r,{\rm eff}}-2p_{t,{\rm eff}}\bigr),\\
R_{\mu\nu}R^{\mu\nu} &= (8\pi)^2\bigl(\rho_{\rm eff}^2+p_{r,{\rm eff}}^2+2p_{t,{\rm eff}}^2\bigr).
\end{aligned}
\label{eq:Ricciinvariants}
\end{equation}
and remain finite because the effective source is finite throughout the compact core. In an orthonormal frame the independent Riemann components can be expressed in terms of $F=1-b/r$, $F'$, $\Phi'$, and $\Phi''$; the resulting Kretschmann scalar and Weyl square are finite at the throat and smooth on the whole manifold.

In an orthonormal frame adapted to Eq.~\eqref{eq:metric}, the independent curvature components can be organized as
\begin{equation*}
\begin{aligned}
\mathcal R_1 &= -F\bigl(\Phi''+\Phi'^2\bigr)-\frac12F'\Phi',\\
\mathcal R_2 &= -\frac{F\Phi'}{r},\\
\mathcal R_3 &= -\frac{F'}{2r},\\
\mathcal R_4 &= \frac{b}{r^3}.
\end{aligned}
\end{equation*}
so that the quadratic invariants take the explicit form
\begin{equation*}
\begin{aligned}
K &= 4\mathcal R_1^2+8\mathcal R_2^2+8\mathcal R_3^2+4\mathcal R_4^2,\\
C^2 &= K-2R_{\mu\nu}R^{\mu\nu}+\frac13R^2.
\end{aligned}
\end{equation*}
Outside the compact core the Ricci sector vanishes identically and the curvature reduces to the exact Schwarzschild values:
\begin{equation}
R=R_{\mu\nu}R^{\mu\nu}=0,
\qquad
K=C^2=\frac{48M^2}{r^6}.
\label{eq:extcurvature}
\end{equation}
The ADM mass on each asymptotic end is therefore
\begin{equation}
M_{\ADM}=M.
\label{eq:ADMmass}
\end{equation}
This highlights an important conceptual point: the compact-support construction localizes the Ricci-curvature support and both the effective and reconstructed matter sectors, but it does not eliminate the exterior Weyl curvature generated by the Schwarzschild mass.

The shell-free nature of the compact boundary follows from the matching conditions. Let $\Sigma_c$ denote the timelike hypersurface $r=\Rc$. Because Eq.~\eqref{eq:flatmatching} forces the metric and all radial derivatives to agree with their Schwarzschild values at $\Sigma_c$, the induced metric and extrinsic curvature coincide on the two sides of the hypersurface. The Darmois--Israel surface tensor therefore vanishes identically,
\begin{equation}
S^{a}{}_{b}=0.
\label{eq:surfzero}
\end{equation}
Equivalently, the full Darmois--Israel tensor may be written as
\begin{equation*}
S^{a}{}_{b}=-\frac{1}{8\pi}\Bigl([K^{a}{}_{b}]-\delta^{a}{}_{b}[K]\Bigr),
\end{equation*}
and every jump vanishes because the induced metric and extrinsic curvature agree with their Schwarzschild values on both sides of $\Sigma_c$. Hence the compact boundary is not merely free of a thin shell in the weak sense of vanishing surface density; it is free of any distributional stress tensor at all.

\begin{figure}[tbp]
\centering
\includegraphics[width=\textwidth]{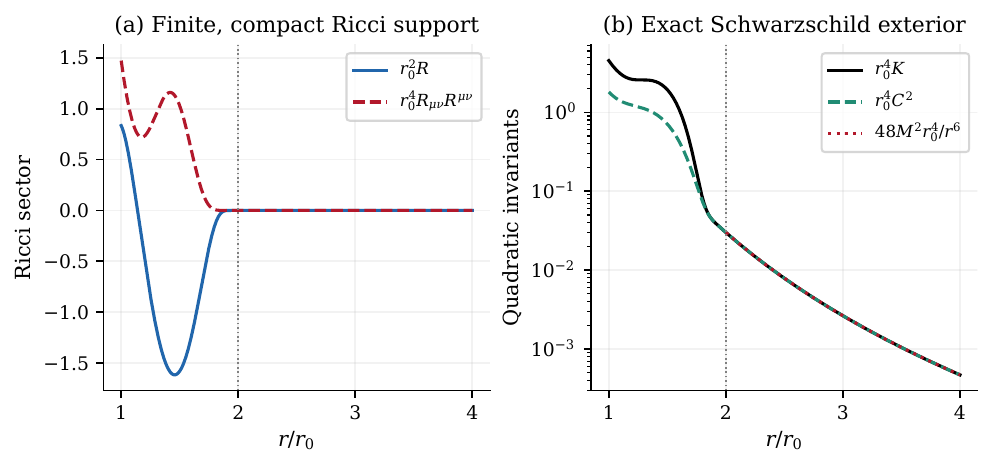}
\caption{Curvature invariants and asymptotic structure for the baseline compact-support wormhole.
Panel (a) shows the Ricci-sector invariants $\hat R=r_0^2R$ and $\widehat{R_{\mu\nu}R^{\mu\nu}}=r_0^4R_{\mu\nu}R^{\mu\nu}$, which remain finite at the throat and vanish identically outside the compact core.
Panel (b) shows the Kretschmann scalar $\hat K=r_0^4K$ and Weyl square $\widehat{C^2}=r_0^4C_{\alpha\beta\gamma\delta}C^{\alpha\beta\gamma\delta}$.
Outside the compact-support region the Ricci sector vanishes, whereas the Weyl and Kretschmann invariants reduce to the exact Schwarzschild tail. The figure summarizes smooth throat behavior, compactly supported Ricci curvature, and exact Schwarzschild vacuum outside the core.}
\label{fig:curvature_regular_asymptotic}
\end{figure}

The invariant statements in this section apply to the geometric seed
independently of reconstruction. On a full regular branch, the implicit
function theorem and the flat effective sources also make the physical
stress tensor smooth and exactly zero outside the core. If a selected root
ends at a fold, only the geometric regularity conclusion persists; a complete
physical matter realization has not been supplied for that parameter value.
The source is conditional on the stated fluid closure, and neither force
balance nor positivity of a scan margin proves dynamical stability.

\section{Reproducibility and Numerical Validation}
\label{sec:reproducibility}
The supplementary package contains the Python geometry and branch solver,
all figure-generation code, the phase grids, benchmark data, and validation
results. It uses NumPy, SciPy, Matplotlib, and mpmath. The principal commands are
\texttt{python code/validate.py} and \texttt{python code/reproduce.py}.
No observational data or fitted parameters enter the calculation.

The cutoff integrals are evaluated directly as tails. For integer $n\ge0$,
the substitution $a=1/(1-x)$ gives
\begin{equation}
\mathcal J_n(x)=\ee\sum_{k=0}^n(-1)^k\binom{n}{k}
\frac{E_{k+2}(a)}{a^{k+1}},\qquad
E_j(a)=\int_1^\infty\frac{\ee^{-at}}{t^j}\dd t.
\label{eq:tail_eval}
\end{equation}
Thus $\delta b=\Delta(\beta\mathcal J_m-s_0\mathcal J_0)$ is computed without
subtracting an almost-vacuum shape function from $2M$.
We also remove the Schwarzschild vacuum terms algebraically before computing
effective pressures. Writing $F=1-b/r$, $\psi=\delta\Phi$, and
$s=\Phi_{\Sch}'$, a useful stable expression is
\begin{equation}
8\pi p_{r,{\rm eff}}=-\frac{\delta b}{r^2(r-2M)}
+\frac{2F}{r}\psi'.
\label{eq:stable_pr}
\end{equation}
The angular expression is evaluated by the same exact subtraction; independent
quadrature tests check Eq.~\eqref{eq:tail_eval}.

The main phase grids start from 1601 uniform $x$ values, and the profile figures
start from 4001. Local discriminant extrema are refined to detect narrow
root-multiplicity intervals between grid points. Ordered real roots are
continued between discriminant events. Every event is bracket-refined before
the surviving root is selected; using a grid midpoint can misidentify the
coalescing pair near a cubic cusp. For a
depressed cubic $y^3+py+q=0$, the sign of $q$ at a double-root event identifies
the coalescing pair. If the selected root belongs to that pair, continuation
stops. A change in the multiplicity of unselected roots does not terminate it.
Benchmark event locations are refined with a bracketing root solver using
absolute $x$ tolerance $2\times10^{-13}$; their values agree to better than
$2\times10^{-10}$ on grids of 401, 801, 1601, and 3201 points.
The 801-point and 1601-point reconstructions give the same phase classification
at 273 common parameter points per map, for 1092 comparisons in total.
Critical tangencies or exceptional nonsimple continuations are outside the
strict admissibility set and require separate analysis.

Weak-source roots are Newton-refined near $\chi=0$ to avoid loss of the GR
limit through cancellation in closed cubic formulas. Cubic residuals and
the original three algebraic field equations are checked on every regular
benchmark. The maximum field-equation residual in hatted variables is below
$10^{-12}$; the supplied data record the actual values and software versions.
The exact $\alpha=0$ limit and its continuous small-coupling approach are tested
separately.

For a finite-difference force-balance check on $0.02<x<0.9$, the regular
baseline $\Pbar$ benchmark gives the convergence in Table~\ref{tab:tov_convergence}.
The second-order decrease is consistent with the differentiation stencil.
The force density is evaluated from the complete Eq.~\eqref{eq:force} for
both matter prescriptions.
\begin{table}[tbp]
\caption{Maximum absolute dimensionless TOV residual for the baseline
$\Pbar$ benchmark at $\alpha=0.5$, using second-order finite differences.}
\label{tab:tov_convergence}
\centering\fontsize{8}{10}\selectfont
\begin{tabular}{rc}\toprule
Grid points & Maximum residual\\\midrule
1001 & $1.31\times10^{-5}$\\
2001 & $3.27\times10^{-6}$\\
4001 & $8.18\times10^{-7}$\\
8001 & $2.05\times10^{-7}$\\
\bottomrule
\end{tabular}

\end{table}

The rapidly flat outer tail requires additional precision. A 90-digit
evaluation of Eq.~\eqref{eq:stable_nec} gives approximately
$-1.64687945\times10^{-18}$ at $x=0.977$ and
$-4.73774598\times10^{-43}$ at $x=0.99$ for the baseline with $r_0=1$.
Ordinary floating-point cancellation can replace these negative values by
an apparent zero. We therefore infer no interior NEC endpoint from such a
zero; the exact baseline sign follows from the argument in
Sec.~\ref{sec:reconstruction}. The proper-distance integral uses $x=t^2$ to
remove the throat square-root singularity before adaptive quadrature.

\section{Conclusion}
\label{sec:conclusion}
Flat cutoff deformations produce explicit $C^\infty$ wormhole geometries with
an exact Schwarzschild exterior, a regular throat, and no surface layer at
the compact boundary. The Ricci-supported nonvacuum region is finite, whereas
the exterior Weyl curvature and ADM mass remain those of Schwarzschild.

Within the assumed anisotropic-fluid closures, physical matter reconstruction
reduces to a cubic equation. The vacuum boundary specifies a locally unique
simple root. Its global continuation must be checked using both the inversion
margin and the selected-root Jacobian. A fold can terminate the simple branch
with $1+\chi$ still positive, and a disconnected real root does not restore
the missing continuation.

The numerical examples establish regular branches with positive and
negative throat density within the displayed families. They do not remove
the radial NEC violation at the throat. In the undeformed-redshift family,
the violation extends through the open core and vanishes at its exact boundary.
The conclusions are conditional existence statements for the stated closures
and ansatz families. A microphysical source model, exceptional branch
continuations, and dynamical perturbation theory remain separate problems.

\section*{Funding}
This research received no external funding.

\section*{Competing Interests}
The authors declare that they have no competing interests.

\section*{Data Availability}
No observational data were used. The numerical source code, phase grids,
benchmark values, and validation results underlying the figures are supplied
as supplementary material.



\clearpage
\begin{thebibliography}{99}

\bibitem{MorrisThorne1988}
M.~S.~Morris and K.~S.~Thorne,
``Wormholes in spacetime and their use for interstellar travel: A tool for teaching general relativity,''
Am.\ J.\ Phys.\ \textbf{56}, 395 (1988).
\url{https://doi.org/10.1119/1.15620}.

\bibitem{Visser1989}
M.~Visser,
``Traversable wormholes from surgically modified Schwarzschild spacetimes,''
Nucl.\ Phys.\ B \textbf{328}, 203 (1989).
\url{https://doi.org/10.1016/0550-3213(89)90100-4}.

\bibitem{VisserBook1995}
M.~Visser,
\emph{Lorentzian Wormholes: From Einstein to Hawking}
(AIP Press, New York, 1995).

\bibitem{HochbergVisser1998}
D.~Hochberg and M.~Visser,
``Null energy condition in dynamic wormholes,''
Phys.
Rev.
Lett.
\textbf{81}, 746 (1998).
\url{https://doi.org/10.1103/PhysRevLett.81.746}.

\bibitem{HochbergMolinaParisVisser1999}
D.~Hochberg, C.~Molina-Paris, and M.~Visser,
``Tolman wormholes violate the strong energy condition,''
Phys.
Rev.
D
\textbf{59}, 044011 (1999).
\url{https://doi.org/10.1103/PhysRevD.59.044011}.

\bibitem{Darmois1927}
G.~Darmois,
\emph{Les equations de la gravitation einsteinienne}
(Memorial des sciences mathematiques, no.~25, 1927).

\bibitem{Israel1966}
W.~Israel,
``Singular hypersurfaces and thin shells in general relativity,''
Nuovo Cimento B \textbf{44}, 1 (1966);
Erratum Nuovo Cimento B \textbf{48}, 463 (1967).

\bibitem{PoissonVisser1995}
E.~Poisson and M.~Visser,
``Thin-shell wormholes: Linearization stability,''
Phys.
Rev.
D
\textbf{52}, 7318 (1995).
\url{https://doi.org/10.1103/PhysRevD.52.7318}.

\bibitem{Eiroa2008}
E.~F.~Eiroa,
``Stability of thin-shell wormholes with spherical symmetry,''
Phys.
Rev.
D
\textbf{78}, 024018 (2008).
\url{https://doi.org/10.1103/PhysRevD.78.024018}.

\bibitem{MazharimousaviHalilsoy2014}
S.~H.~Mazharimousavi and M.~Halilsoy,
``Flare-out conditions in static thin-shell wormholes,''
Phys.
Rev.
D
\textbf{90}, 087501 (2014).
\url{https://doi.org/10.1103/PhysRevD.90.087501}.

\bibitem{Harko2011}
T.~Harko, F.~S.~N.~Lobo, S.~Nojiri, and S.~D.~Odintsov,
``$f(R,T)$ gravity,''
Phys.\ Rev.\ D \textbf{84}, 024020 (2011).
\url{https://doi.org/10.1103/PhysRevD.84.024020}.

\bibitem{BanerjeeTangphatiPradhan2023}
T.~Tangphati, A.~Banerjee, and A.~Pradhan,
``Wormholes and energy conditions in $f(R,T)$ gravity,''
Int.\ J.\ Geom.\ Methods Mod.\ Phys.\ \textbf{21}, 2450109 (2024).
\url{https://doi.org/10.1142/S0219887824501093}.

\bibitem{MoraesSahoo2017}
P.~H.~R.~S.~Moraes and P.~K.~Sahoo,
``Modeling wormholes in $f(R,T)$ gravity,''
Phys.\ Rev.\ D \textbf{96}, 044038 (2017).
\url{https://doi.org/10.1103/PhysRevD.96.044038}.

\bibitem{MoraesSahoo2018}
P.~K.~Sahoo, P.~H.~R.~S.~Moraes, and P.~Sahoo,
``Wormholes in $R^2$-gravity within the $f(R,T)$ formalism,''
Eur.\ Phys.\ J.\ C \textbf{78}, 46 (2018).
\url{https://doi.org/10.1140/epjc/s10052-018-5538-1}.

\bibitem{ChandaDeyPaul2021}
A.~Chanda, S.~Dey, and B.~C.~Paul,
``Morris--Thorne wormholes in modified $f(R,T)$ gravity,''
Gen.
Relativ.
Gravit.
\textbf{53}, 78 (2021).
\url{https://doi.org/10.1007/s10714-021-02847-7}.

\bibitem{MoraesSahoo2019}
P.~H.~R.~S.~Moraes and P.~K.~Sahoo,
``Wormholes in exponential $f(R,T)$ gravity,''
Eur.
Phys.
J.
C
\textbf{79}, 677 (2019).
\url{https://doi.org/10.1140/epjc/s10052-019-7206-5}.

\bibitem{MishraSharmaDubeyPradhan2020}
A.~K.~Mishra, U.~K.~Sharma, V.~C.~Dubey, and A.~Pradhan,
``Traversable wormholes in $f(R,T)$ gravity,''
Astrophys.
Space.
Sci.
\textbf{365}, 34 (2020).
\url{https://doi.org/10.1007/s10509-020-3743-5}.

\bibitem{KiroriwalKumarMauryaChaudhary2024}
S.~Kiroriwal, J.~Kumar, S.~K.~Maurya, and S.~Chaudhary,
``A comparative study of wormhole geometries under two different modified gravity formalism,''
Eur.
Phys.
J.
C
\textbf{84}, 414 (2024).
\url{https://doi.org/10.1140/epjc/s10052-024-12744-3}.

\bibitem{MoraesSahooQuadratic2018}
P.~H.~R.~S.~Moraes and P.~K.~Sahoo,
``Nonexotic matter wormholes in a trace of the energy-momentum tensor squared gravity,''
Phys.\ Rev.\ D \textbf{97}, 024007 (2018).
\url{https://doi.org/10.1103/PhysRevD.97.024007}.

\bibitem{LoboOliveira2009}
F.~S.~N.~Lobo and M.~A.~Oliveira,
``Wormhole geometries in $f(R)$ modified theories of gravity,''
Phys.\ Rev.\ D \textbf{80}, 104012 (2009).
\url{https://doi.org/10.1103/PhysRevD.80.104012}.

\bibitem{BarrientosRubilar2014}
J.~Barrientos and G.~F.~Rubilar,
``Comment on `$f(R,T)$ gravity',''
Phys.\ Rev.\ D \textbf{90}, 028501 (2014).
\url{https://doi.org/10.1103/PhysRevD.90.028501}.

\bibitem{FisherCarlson2019}
S.~B.~Fisher and E.~D.~Carlson,
``Reexamining $f(R,T)$ gravity,''
Phys.\ Rev.\ D \textbf{100}, 064059 (2019).
\url{https://doi.org/10.1103/PhysRevD.100.064059}.

\bibitem{HarkoMoraes2020}
T.~Harko and P.~H.~R.~S.~Moraes,
``Comment on `Reexamining $f(R,T)$ gravity',''
Phys.\ Rev.\ D \textbf{101}, 108501 (2020).
\url{https://doi.org/10.1103/PhysRevD.101.108501}.

\bibitem{FisherCarlson2020}
S.~B.~Fisher and E.~D.~Carlson,
``Reply to `Comment on ``Reexamining $f(R,T)$ gravity''',''
Phys.\ Rev.\ D \textbf{101}, 108502 (2020).
\url{https://doi.org/10.1103/PhysRevD.101.108502}.
\end{thebibliography}
\end{document}